\documentclass[twocolumn,preprintnumbers,amsmath,amssymb,floatfix,superscriptaddress,nofootinbib]{revtex4}

\usepackage{amsmath,hyperref,amssymb}
\usepackage{graphicx}
\usepackage{dcolumn}
\usepackage{bm}
\usepackage{verbatim}
\usepackage{tabularx}
\usepackage{slashed}
\usepackage{cancel}
\usepackage[boxsize=2em,aligntableaux=center]{ytableau}
\usepackage{float}
\usepackage{color}

\def\be{\begin{equation}}
\def\ee{\end{equation}}
\def\bea{\begin{eqnarray}}
\def\eea{\end{eqnarray}}

\newcommand{\LL}{\mathcal{L}}

\begin{document}

\vspace*{-30mm}

\title{Axion interferometry}

\author{William DeRocco}
\affiliation{Stanford Institute for Theoretical Physics, \\
Stanford University, Stanford, CA 94305, USA}
\author{Anson Hook}
\affiliation{Stanford Institute for Theoretical Physics, \\
Stanford University, Stanford, CA 94305, USA}
\affiliation{Maryland Center for Fundamental Physics, Department of Physics\\
University of Maryland, College Park, MD 20742.}

\vspace*{1cm}

\begin{abstract} 

We propose using interferometry of circularly polarized light as a mechanism by which to test for axion dark matter.
These interferometers differ from standard interferometers only by the addition of a few quarter waveplates to preserve the polarization of light upon reflection.  We show that using current technology, interferometers can probe new regions of axion parameter
space up to a couple orders of magnitude beyond current constraints.

\end{abstract}

\maketitle

\section{Introduction}

One of the leading candidates for dark matter (DM) is a light pseudo-scalar derivatively coupled to the Standard Model (SM).
The most well-known example of such a candidate is the QCD axion~\cite{Peccei:1977hh,Peccei:1977ur,Weinberg:1977ma,Wilczek:1977pj}.
The axion can have a multitude of different couplings to the SM.  The coupling that produces the effect of interest in this article is
\bea
\label{Eq: coupling}
\LL \supset \frac{a}{4 f} F \tilde F
\eea
which is the axion coupling to photons.  While in the simplest models of the QCD axion, the axion-photon coupling is a function of the axion mass, there exist models where the coupling to photons is a free parameter (i.e. $f$ is independent of $m_a$)~\cite{Farina:2016tgd,Agrawal:2017cmd}.  We consider axions, which do not necessarily have to be the QCD axion, where $f$ and $m_a$ are independent of each other.  These generalized axions are sometimes called axion-like particles (ALPs).  There are many proposals for experiments to look for axions and ALPs.  See Refs.~\cite{Horns:2012jf,Budker:2013hfa,Graham:2013gfa,Stadnik:2014tta,Hill:2015vma,Kahn:2016aff,Abel:2017rtm} for a small subset of these proposals.

In the presence of ALP dark matter, the coupling shown in Eq.~\ref{Eq: coupling} generates new terms in Maxwell's equations. In vacuum, the equations become
\begin{align}
&\nabla \cdot \vec{E} = -\frac{1}{f}\nabla a \cdot\vec{B}& \\
&\nabla \times \vec{E} = -\frac{\partial \vec{B}}{\partial t}& \\
&\nabla \cdot \vec{B} = 0 &\\
&\nabla \times \vec{B} = \frac{\partial \vec{E}}{\partial t} + \frac{1}{f}(\dot{a}\vec{B} + \nabla a \times \vec{E})&
\end{align}
Turning Maxwell's equations into the wave equation for light and taking the limit of a light non-relativistic axion ($v \ll 1$ and $m_a \ll \omega$), one arrives at the relation
\bea
\frac{\partial^2 \vec{E}}{\partial t^2} - \nabla^2 \vec{E} = \frac{\dot{a}}{f}(\vec{\nabla} \times \vec{E})
\eea
Substituting a plane-wave solution yields a modified dispersion relation:
\bea
-\omega^2 + k^2 \mp \frac{\dot{a}}{f}k = 0
\eea
This is just the well-known effect that the presence of ALP dark matter causes a difference in phase velocity between right and left circularly polarized light.
This effect is often equivalently stated as the fact that a background axion field causes the polarization angle of linearly polarized light to slowly rotate.
It follows that the phase velocity of left and right polarized light is 
\bea
v_\text{phase} \approx 1 \pm \frac{\dot a}{2 k f}
\eea

As the effect of axion dark matter is to change the phase velocity of circularly polarized light, the natural experiment to build is an 
interferometer where one arm has left circularly polarized light while the other arm has right polarized light. Axion DM would produce a difference in phase velocity between the two arms, generating an interference pattern.

\section{Mapping between gravitational waves and axion DM}

If the light in the interferometer is circularly polarized, there is an exact mapping between the effects of axions and gravitational waves.
Therefore all of the literature on gravitational wave interferometry can be imported directly into axion interferometry.

\begin{figure}
  \centering
  \includegraphics[width=0.45\textwidth]{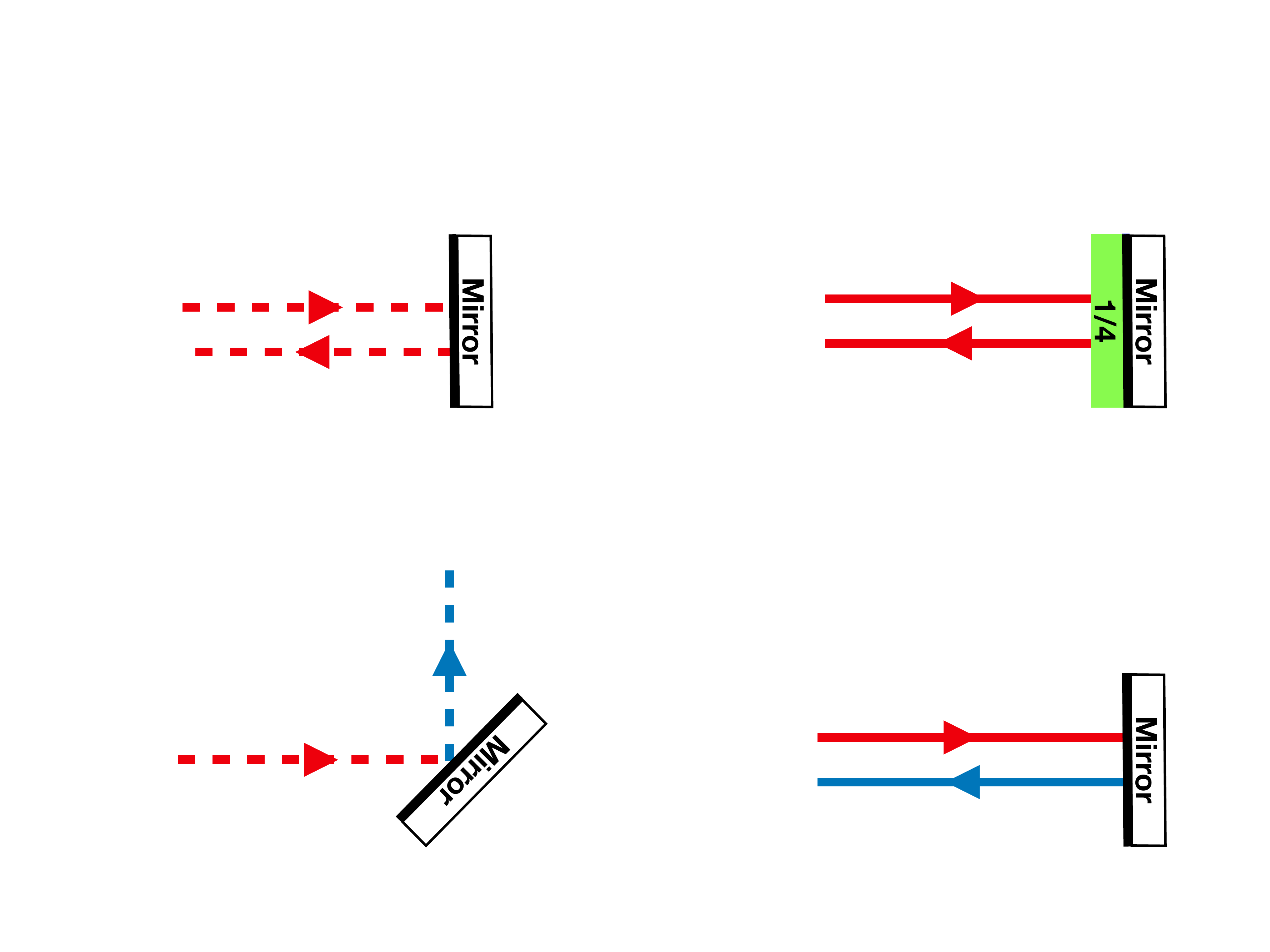}
  \caption{ How to map reflections off of mirrors between axion interferometers and their equivalent gravity wave interferometers.  On the left are the gravity wave interferometer set-ups (dashed lines) and on the right are the axion interferometer set-ups (solid lines).  Red lines indicate light going in the $x$ direction or $\circlearrowleft$ polarized light and blue lines indicate light going in the $y$ direction or $\circlearrowright$ polarized light. The comparison is drawn between the $x$/$y$ direction of light for GWs and left/right circular polarizations for axion DM because the effect of a GW is a change in the path length between the $x$ and $y$ directions while the effect of axion DM is a change in the path length between right and left circular polarizations.
    \label{Fig: equivalent}}
\end{figure}

To map between gravitational waves and axions, we compare an axion interferometer with left and right polarized light respectively in each of the two arms with a gravitational wave interferometer with arms along the $x$ and $y$ directions subject to a + polarized gravitational wave propagating along the $z$-axis. Since the velocity of dark matter is small ($v \sim 10^{-3}$), the length of the interferometer is $\ll 1/m_a v$, so it is safe to neglect the effect of the spatial gradients of the axion field.  The equivalent GW propagates along the $z$-axis because this maps to the situation of having negligible spatial gradients in the axion DM.

To map between the amplitude of the gravitational wave $h_0$ and the effect of the axion DM, we note that the axion field behaves as a classical field due to its large occupation number.  The axion field is approximately
\bea
a(t) = a_0 \cos(m_a t + k_a z) .
\eea
Using the dispersion relation and neglecting spatial gradients, this gives us an effective path length of 
\bea
L_{\circlearrowleft,\circlearrowright} = \int^{t_0 + \tau}_{t_0} 1 \pm \frac{m_a a_0}{2 f \omega}\cos(m_a t)~dt
\eea

Comparison to the standard formula for path length in the case of gravitational waves~\cite{Maggiore:1900zz}
\bea 
L_{x,y} = \int^{t_0 + \tau}_{t_0} 1 \pm \frac{1}{2}h_0\cos(\omega_g t)~dt
\eea
shows that the correct mapping between the two scenarios is
\bea 
\label{Eq: h0}
h_0 \rightarrow \frac{m_a a_0}{f \omega} = \frac{\sqrt{2 \rho_\text{DM}}}{\omega f} \qquad \omega_g \rightarrow m_a
\eea
where we have used that the local dark matter density $\rho_\text{DM} = \frac{1}{2}(m_a a_0)^2 \sim 0.3$ GeV/cm$^3$. $\omega$ denotes the angular frequency of the laser light while $\omega_g$ denotes the angular frequency of the gravitational wave.
Finally, since axion DM is constantly streaming through us with a quality factor $Q \sim \frac{1}{v^2} \sim 10^6$, it is equivalent to a continuous gravitational wave with similar quality factor.  

It should be noted that unlike a gravitational wave detector, the two arms of the interferometer need not be perpendicular. In fact, they could be run parallel such that the right-handed and left-handed cavities are actually formed by the same mirror. This design could potentially allow for significant reductions in radiation pressure noise (discussed further below).  Power recycling and other improvements that are independent of the arms of the interferometer are mapped between set-ups with no change.

Up until now, we have neglected to discuss a crucial point that reflection off of a mirror inverts the polarization of the laser beam.  Since the axion DM-induced effect is polarization-dependent, the sign of the effect changes upon reflection off of a mirror.  If this effect is to be prevented, polarization-preserving mirrors must be used.  These can be manufactured by adding a quarter-waveplate in front of a mirror or by including a coating on the mirror that produces the same effect.  The mapping between mirrors that perform an equivalent function in a gravitational wave interferometer and an axion interferometer is shown in Fig.~\ref{Fig: equivalent}.

\section{An axion interferometer}

\begin{figure}
  \centering
  \includegraphics[width=0.45\textwidth]{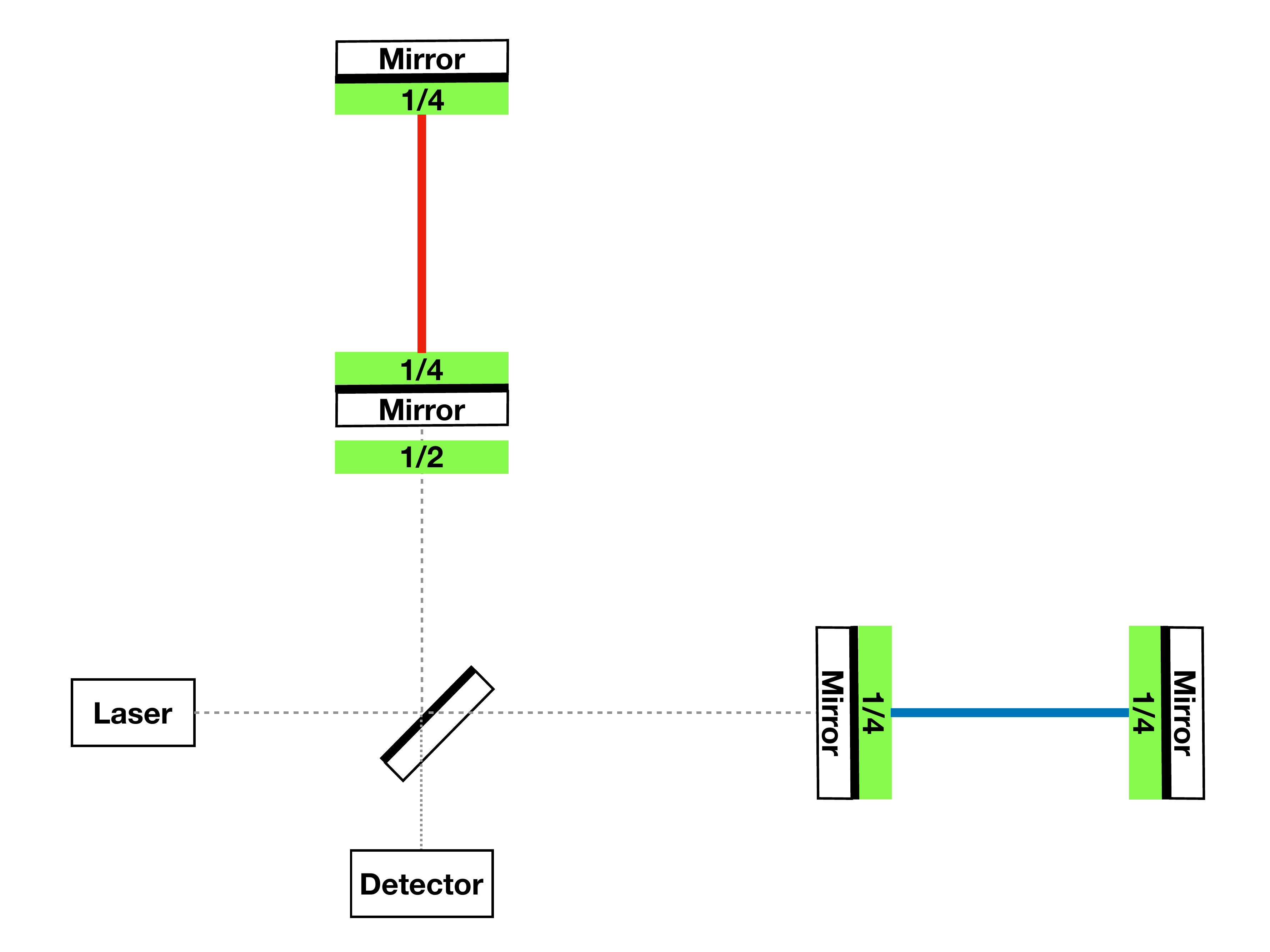}
  \caption{ Diagram of our proposed axion interferometer.  It is a standard Michelson interferometer with the additional waveplates  necessary to preserve polarization. The two arms of the interferometer both consist of Fabry-Perot cavities that allow the effective path length to be higher than a single-traverse interferometer by roughly a factor of the cavity finesse.  The dotted line is linearly polarized light, the red line is $\circlearrowleft$ polarized light and the blue line is $\circlearrowright$ polarized light.  Four quarter waveplates and a half waveplate are used to maintain the circular polarization of the light.
    \label{Fig: interferometer}}
\end{figure}

We are finally in a position to describe our axion interferometer.  It is shown diagrammatically in Fig.~\ref{Fig: interferometer}.
As we do not know the mass of ALP dark matter, we wish to design a broadband detector rather than a resonant detector.  We thus choose our equivalent gravitational wave interferometer to be a typical Michelson interferometer such as those used in experiments such as LIGO~\cite{Abbott:2007kv} and the Holometer~\cite{Chou:2016cye}. 

The proposed experiment is just a standard Michelson interferometer with the addition of four quarter-waveplates and a half-waveplate. Since most beam-splitters require linearly polarized light, the setup is designed in such a way that light only becomes circularly polarized upon entering the cavity due to passing through the first quarter-waveplate. Within the cavity, its polarization is preserved. Upon exiting, the light is reconverted into linearly polarized light by a single pass through the quarter waveplate.

\begin{figure}
  \centering
  \includegraphics[width=0.45\textwidth]{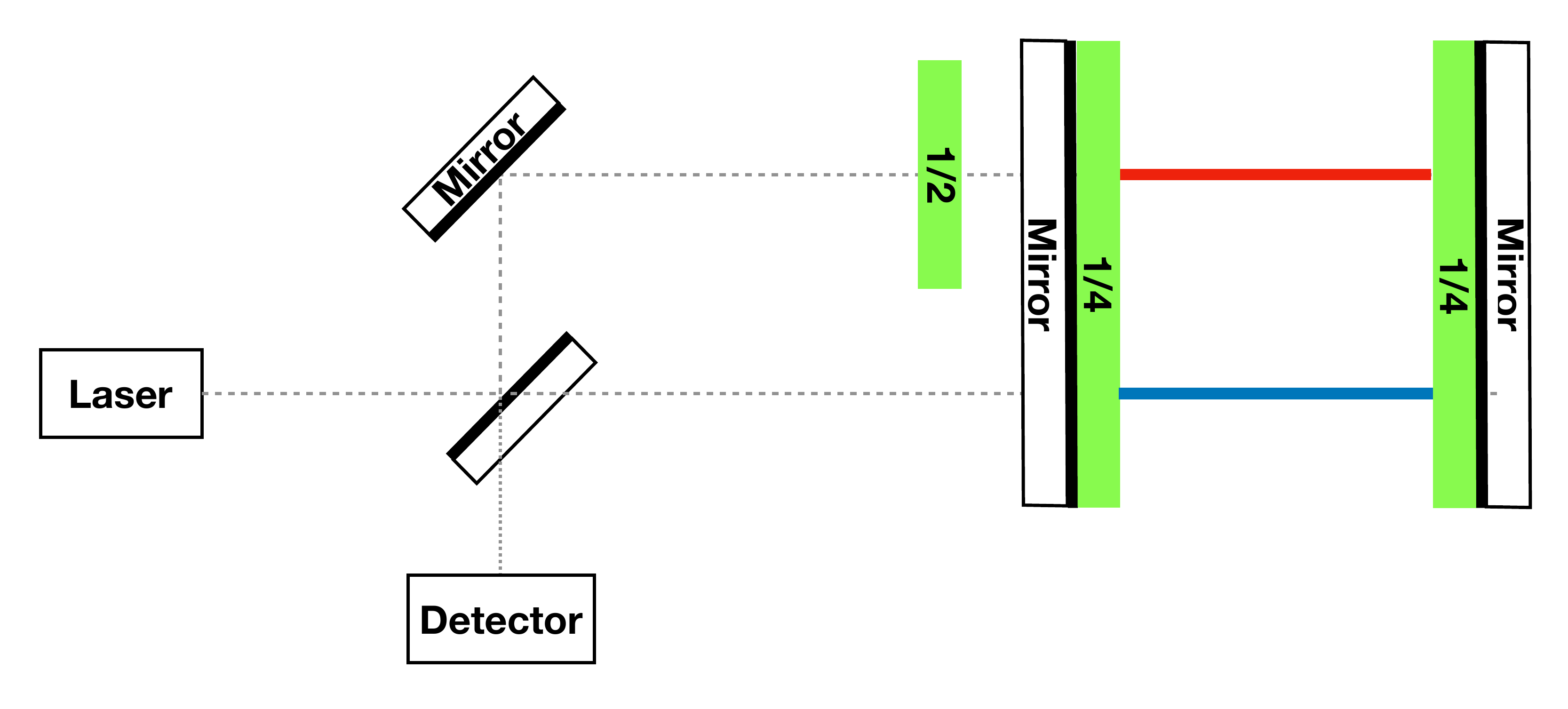}
  \caption{ A diagram of our proposed axion interferometer where the same mirrors are used to form both cavities. The dotted line is linearly polarized light, the red line is $\circlearrowleft$ polarized light and the blue line is $\circlearrowright$ polarized light.  Two quarter waveplates and a half waveplate are used to maintain the circular polarization of the light. This setup cancels the radiation pressure noise associated with the displacement of the mirror, leaving only noise due to radiation torque. Torque noise in this setup can be several orders of magnitude smaller than the radiation pressure  noise experienced by the setup in Fig.~\ref{Fig: interferometer}. 
    \label{Fig: interferometer2}}
\end{figure}

The half-waveplate is included to change the polarization angle of the incident linearly polarized light. Changing the polarization angle of the $y$-oriented light in Fig.~\ref{Fig: interferometer} by $\frac{\pi}{2}$ causes it to be converted into circularly polarized light of opposite handedness as the $x$-oriented cavity. Therefore the two arms of the interferometer feel opposite-sign effects from axion DM, causing interference when the beams are recombined.  As mentioned before, the arms do not need to be perpendicular to each other and could be run using the same mirrors for both cavities to reduce noise.  This improved version of the interferometer is shown diagrammatically in Fig.~\ref{Fig: interferometer2}.

Since our proposed experiment requires the addition of various waveplates, the waveplates must be assessed for potential sources of systematic error. One effect is that the waveplates are not
perfect. Losses in the waveplates and increased thermal noise due to absorption will likely limit the highest possible finesse achievable within a cavity. As such, we choose to display the reach of axion interferometers using finesses of both the easily realizable $10^2$ and the much more speculative $10^6$, which is the highest finesse that current cavities can attain in the absence of any waveplates~\cite{finesse}.

Another possible source of noise is due to birefringent effects coming from reflecting off of these polarization-preserving mirrors. Previous experiments have mainly focused on controlling birefringent effects in the context of linearly polarized light~\cite{DellaValle:2013dwa,Cadene:2013bva}.  It will be an experimental question whether or not these effects can be sufficiently suppressed as to be a subdominant source of noise.

\section{Parameter space probed by axion interferometers}

\begin{figure}
  \centering
  \includegraphics[width=0.45\textwidth]{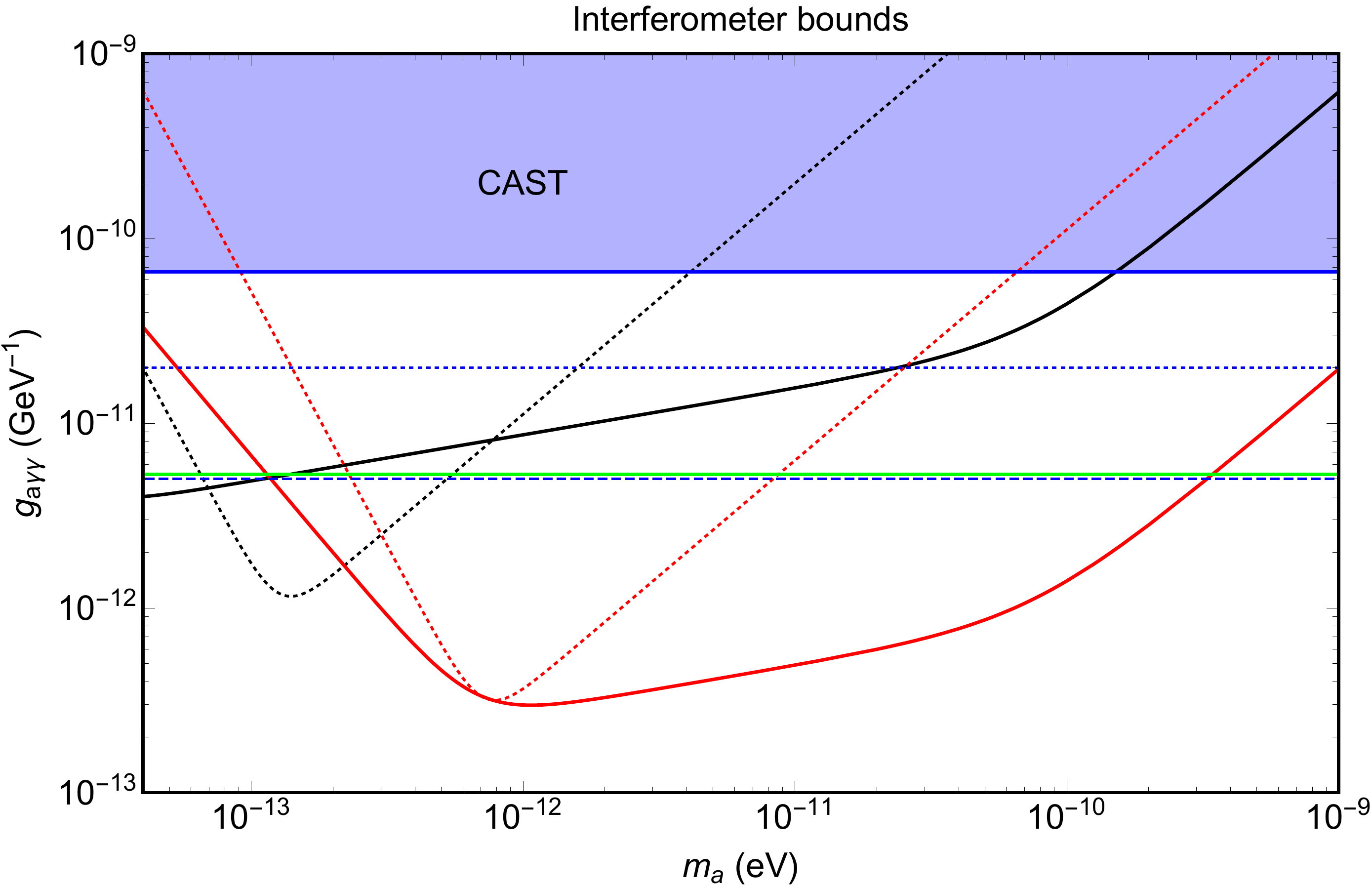}
  \caption{The reach of an axion interferometer in $g_{a\gamma\gamma} = 1/f$ as a function of mass. We cut off the plot at frequencies of roughly 10 Hz where there start to be unavoidable sources of noise stemming from gravity gradient and seismic noise.  The plot was made assuming a 40 m long interferometer and 10 kg mirrors.  The solid (dotted) line shows $\mathcal{F} = 10^2$ ($\mathcal{F} = 10^6$). The black (red) line assumes a power of 1 kW (1 MW) circulating inside the Fabry-Perot cavities.  Bounds placed by CAST are shown in blue~\cite{Anastassopoulos:2017ftl}.  Constraints coming from the production of axions in supernova and subsequent conversions into photons in the interstellar medium are shown in green~\cite{Payez:2014xsa}.  The reach of the other proposed experiments IAXO~\cite{Irastorza:2013dav} (ALPS II~\cite{Bahre:2013ywa}) are shown in dashed (dotted) blue.
    \label{Fig: bound40}}
\end{figure}

In this section, we calculate the reach of an axion interferometer assuming that noise from the waveplates has been mitigated such that we are at the standard quantum limit (SQL) as is the case in LIGO and the Holometer for a range of frequencies. Under this assumption, the data analysis is identical to that of a continuous gravitational wave detector.  The standard SQL signal-to-noise ratio (SNR)~\cite{Maggiore:1900zz} is
\bea
\label{Eq: SNR}
\text{SNR} = \frac{h_0}{S_{SQL}^{1/2}}  \left( T \tau \right)^{\frac{1}{4}}
\eea
where $T$ is the observation time, $\tau$ is the coherence time of the axion field ($=\frac{2\pi}{m_a v^2}$) and $h_0$ is given by Eq.~\ref{Eq: h0}\footnote{Stochastic backgrounds are usually searched for by looking for correlations in the output power of multiple detectors.  Unlike most stochastic backgrounds, the axion has a large coherence time.  A version of the usual search modified to apply to large $Q$ signals would give similar sensitivity to our matched waveform approach.}.
The $T^{\frac{1}{4}}$ dependence is due to the fact that the axion field is only coherent on a timescale $\tau \sim (m_a v^2)^{-1}$, so the sensitivity of the experiment increases as $\sqrt{T}$ up until the coherence time, then as $T^{\frac{1}{4}}$.  

The SQL is a combination of shot noise and radiation pressure noise, $S_{SQL} = S_\text{shot} + S_\text{radiation}$.  The shot noise is
\bea
S_\text{shot}^{1/2} = \frac{1}{4L} \sqrt{\frac{2 \lambda}{\pi P_0}} \frac{\sin \phi_0}{\sin 2 \phi_0} \sqrt{1 + r^2 - 2 r \cos 2 m_a L}
\eea
where $P_0$ is the power incident on the beam-splitter, $\lambda$ is the wavelength of laser light, $\phi_0$ is how far off of the dark spot the interferometer is tuned to, $L$ is the length of the cavity, and $r$ is the reflectivity of the mirror closer to the beam-splitter (the reflectivity of the further mirror in a cavity is taken to be 1).  The radiation pressure noise is
\bea
S_\text{radiation} = \frac{16 \mathcal{F}}{M L m_a^2} \sqrt{\frac{P}{\pi \lambda}} \frac{m_a L}{\sin m_a L} \frac{1-r^2}{2 \sqrt{1 + r^2 - 2 r \cos 2 m_a L}} \nonumber
\eea
where M is the mass of the mirror and $\mathcal{F}$ is the finesse of a cavity ($r \approx 1-\frac{\pi}{\mathcal{F}}$). 

This noise can be reduced by running the interferometer in the configuration shown in Fig.~\ref{Fig: interferometer2}. Since both cavities are now formed by the same mirrors, any change in the displacement of the mirror occurs equally in both cavities, hence the overall displacement noise due to radiation pressure is cancelled. What remains is radiation torque noise, which arises when fluctuations in power between the two beams cause a torque on the mirror, leading to slightly different path-lengths for the two beams. This noise is then given by
\bea
S_\text{radiation torque} = \frac{M r^2}{I} S_\text{radiation}
\eea
where $r$ is the distance between a beam and the center of a mirror and $I$ is the moment of inertia of the mirror.  By reducing $r$, the noise from radiation torque can be made to be several orders of magnitude smaller than the usual radiation pressure noise.

\begin{figure}
  \centering
  \includegraphics[width=0.45\textwidth]{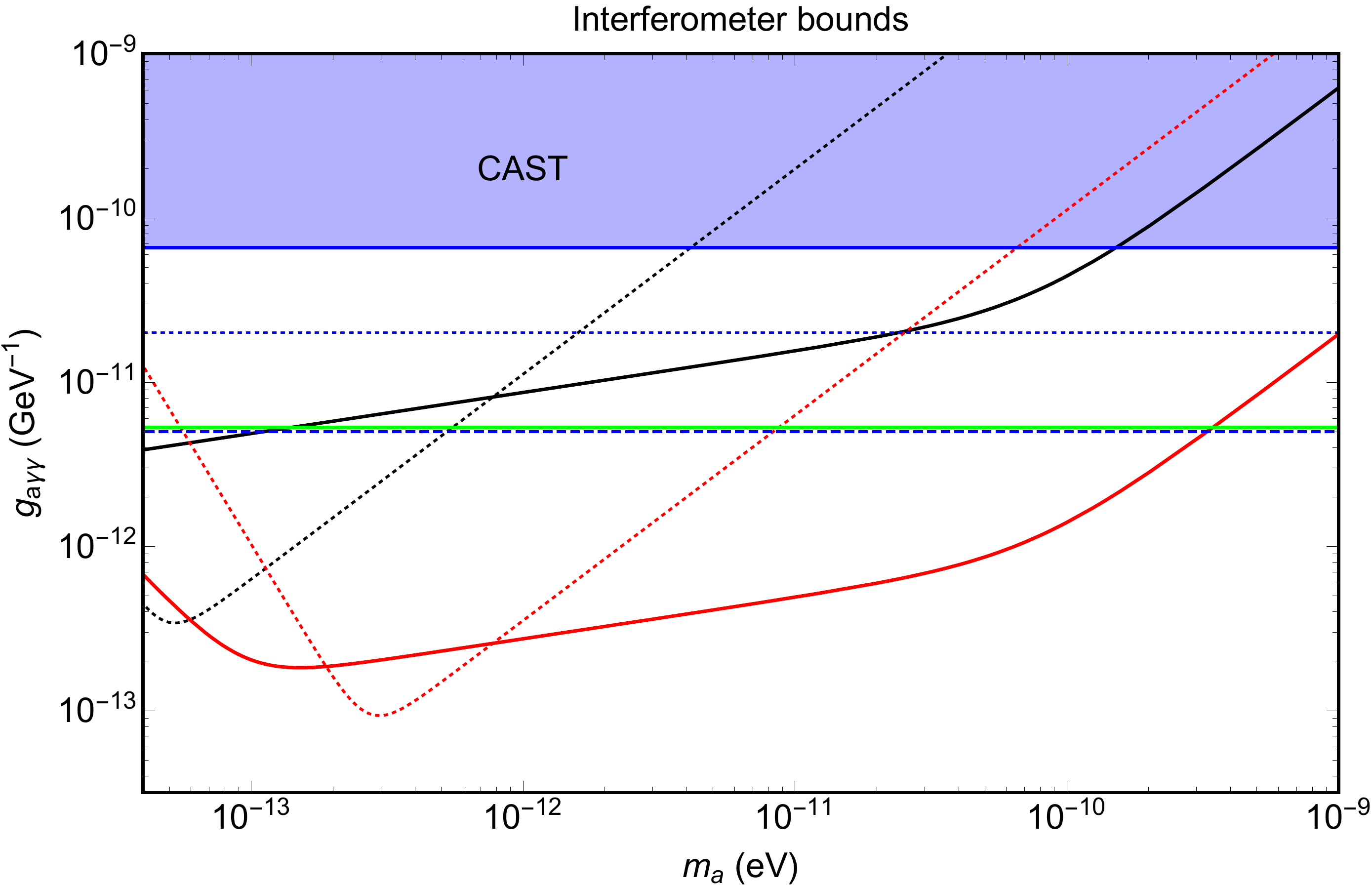}
  \caption{Same as Fig.~\ref{Fig: bound40} but using the configuration shown in Fig.~\ref{Fig: interferometer2}.  Radiation pressure noise is cancelled leaving only radiation torque noise.  We take the beams to be separated by 1 cm and the mirror to be circular and 10 cm in diameter.
    \label{Fig: bound40T}}
\end{figure}

To compute the reach shown in Fig.~\ref{Fig: bound40} and Fig.~\ref{Fig: bound40T}, we set SNR  $=1$ and solve for $f$ as a function of $m_a$.  The dominant experimental constraint is the power contained within the cavity, which is given by $(\frac{2}{\pi}) P_0 \mathcal{F}$. The incident power and finesse must be chosen such that this quantity does not exceed several hundreds of kW, which is the maximal power that can be currently contained within a cavity~\cite{finesse}. For this reason, one cannot increase $P_0$ arbitrarily without a corresponding reduction in finesse.  

 Fig.~\ref{Fig: bound40} (Fig.~\ref{Fig: bound40T}) was made taking $L = 40$ m using the design shown in Fig.~\ref{Fig: interferometer} (Fig.~\ref{Fig: interferometer2}).  Solid (dotted) lines have a finesse of $10^2$ ($10^6$).  We took a standard 1064 nm laser, $\phi_0 = \pi/4$, $M = 10$ kg, and $T = 30$ days.  As one of the limiting factors is the power stored in the cavity, we show exclusions in black (red) using the easily-accessible (more difficult) value of 1 kW (1 MW) of power stored in the cavity.

The general shape of the reach curves can be understood as follows.  At low frequencies, the reach curves weaken due to radiation pressure noise.  At high frequencies,  a given reach curve has two different slopes in different regimes of the axion mass.  The first, more gradual weakening of the reach curve comes from the change in the coherence time as the mass increases.  The second, steeper slope occurs when the axion field is fluctuating on time-scales comparable to or shorter than the trapping time of the cavity.  The phase shift begins to be averaged out since the light is trapped for greater than one half-period of the axion field.  A longer trapping time (equivalently a longer effective arm length) therefore means that the interferometer starts losing sensitivity at higher axion masses.

An interesting aspect of this experimental design is that interferometers with larger effective arm length do not necessarily probe more of parameter space than interferometers with smaller effective arm length. As can be seen from the figures, interferometers with different finesses probe different regions of parameter space.  The reason for this difference is that,  as mentioned before, larger finesse cavities require lower power input lasers.  Lower power on the beam splitter results in larger noise that can degrade sensitivity.  Therefore axion interferometers of different finesses and laser powers can complement each other to better cover parameter space. Note that while the interferometer with $\mathcal{F} = 10^6$ appears to cover less parameter space than $\mathcal{F} = 10^2$, we have chosen to display it both to contrast our experiment with other axion-detection experiments that often seek a quality factor of $10^6$ and to demonstrate that existing interferometers with low-power lasers could still be repurposed to probe interesting regions of parameter space.

It is worth noting that unlike a gravitational wave detector, the reach of an axion interferometer improves for decreasing $\omega$. This is due to the inverse $\omega$-dependence of $h_0$, which is not present in the case of gravitational waves.  Though the fact that longer wavelengths of light are preferred might suggest that the experiment should attempt to use the longest wavelengths possible, the assumption of shot noise limitation is no longer valid for wavelengths much longer than those of visible light due to the inability to detect single low energy photons. This makes experimental control of noise significantly more difficult at longer wavelengths and weakens the potential sensitivity.  Optimistically, if future advances in Transition Edge Sensors~\cite{TES} and/or Microwave Kinetic Inductance Devices~\cite{MKID} allow for the use of a meV scale standard quantum limited maser, then the reach would be improved by a factor of $\sim 30$.

\section{Conclusion}

In this article, we proposed an interferometer-based search strategy for ALP dark matter.  Because there is a direct mapping between gravitational wave interferometers and axion interferometers, much of the technology developed for interferometry applies equally well to axion detection.  The only technical difference is the addition of quarter waveplates to preserve the polarization of the light.  If an experiment of this sort were to be undertaken, it would be able to push beyond current constraints on ALPs by several orders of magnitude for reasonable regions of parameter space.  Once the ALP mass is known, other designs such as resonant gravity wave interferometers~\cite{Meers:1988wp,Vinet:1988ch} could be transformed into axion interferometers and used as well.

\section*{Acknowledgements}

W.D. and A.H. are supported by DOE Grant DE- SC0012012.  A.H. is also supported by NSF Grant PHY-1620074 and by the Maryland Center for Fundamental Physics (MCFP).  W.D and A.H. thank Gustavo Marques Tavares for collaboration in the early stages of the project.  W.D. and A.H. thank Junwu Huang for useful comments on the manuscript.

\bibliography{ref}

\end{document}